\documentclass[10pt,letterpaper]{article}
\usepackage[top=0.85in,left=0.75in,footskip=0.75in,marginparwidth=1in]{geometry}

\usepackage[right]{lineno}

\raggedright
\usepackage{changepage}

\usepackage[aboveskip=1pt,labelfont=bf,labelsep=period,singlelinecheck=off]{caption}

\makeatletter
\renewcommand{\@biblabel}[1]{\quad#1.}
\makeatother

\usepackage{lastpage,fancyhdr,graphicx}
\usepackage{epstopdf}
\fancyheadoffset[L]{2.25in}
\fancyfootoffset[L]{2.25in}

\usepackage{color}

\definecolor{Gray}{gray}{.25}

\usepackage{graphicx}

\usepackage{sidecap}

\usepackage{wrapfig}
\usepackage[pscoord]{eso-pic}
\usepackage[fulladjust]{marginnote}
\reversemarginpar
\usepackage{stackrel}

\begin{document}
\vspace*{0.35in}


\begin{flushleft}
{\Large
\textbf\newline{More than trees, do we need a complex perspective for sustainable forest management?}
}
\newline

{Oliver L\'opez-Corona}\textsuperscript{1,2},
{Elvia Ram\'irez-Carrillo}\textsuperscript{3},
{Vanessa P\'erez-Cirera}\textsuperscript{1},
{Fernando de Le\'on-Gonz\'alez}\textsuperscript{4},
{Rodolfo Dirzo}\textsuperscript{5},

\bigskip
\bf{1} {Instituto de Investigaci\'on para el Desarrollo con Equidad (EQUIDE), Universidad Iberoamericana, Ciudad de M\'exico, M\'exico.}
\newline
\bf{2} {Centro de Ciencias de la Complejidad (C3), Universidad Nacional Aut\'onoma de M\'exico,Ciudad de M\'exico, M\'exico. }
\newline
\bf{3}{Doctorado en Ciencias Agropecuarias, Universidad Aut\'onoma Metropolitana-Xochimilco, Ciudad de M\'exico, M\'exico}
\newline
\bf{4}{Departamento de Producci\'on Agr\'icola y Animal, Universidad Aut\'onoma Metropolitana-Xochimilco, Ciudad de M\'exico, M\'exico}
\newline
\bf{5}{Department of Biology, Stanford University, Stanford, CA 94305, USA.}
\bigskip
* lopezoliverx@otrasenda.org
\end{flushleft}


\begin{abstract}

Forests are complex systems, and it is necessary to include this characteristic in every forest definition, in order to consider the restriction that this imposes in terms of prediction and control. This lost of predictability and controllability should be incorporated in every Environmental Impact Assessment or management program. We present two case-studies located in Mexico and one in the US to illustrate  three relevant indicators of complexity.  First, we introduce an informational framework to measure the Zoquiapan forest systemic complexity. Then, we analyze complexity changes among different types of forest and management systems, related with  spatial distributions, using data from a floristic study in the Montes Azules National Park. Finally, we analyze time series of $CO_{2}$ fluctuations taken from AMERIFLUX data bases. Our results show firstly that it is possible to measure the systemic complexity of different forests, characteized by a criticality state (1/f noise) which has been proposed as a finger print of complexity. And secondly, that this characteristic can be  used as a proxy of their state of conservation, where the lowest complexity values are found in perturbed areas showing the relevance of the concept and its measurement for forest conservation and management. 

\end{abstract}

\section{Introduction}

There are at least 800 different definitions of forest, some of them are used simultaneously in a same country for different purposes or scales \cite{Lund-2008}. This is in part due to the fact that forest types differ widely, depending on factors such as latitude, climate patterns, soil properties and human interactions. It also depends on who is defining it, an economist might define a forest in a very different way from a forester or farmer, in correspondence with their particular focus and interests.

One of the most widely used definition is that by FAO (1998) which poses that a forest is land area of more than 0.5 ha, with a tree canopy cover of more than 10\%, which is not primarily under agricultural or other specific non-forest land use. In the case of young forests or regions where tree growth is climatically suppressed, the trees should be capable of reaching a height of 5 m \textit{in situ}, and of meeting the canopy cover requirement.

In general, forest definitions are based in two different perspectives. One, associated with quantitative cover/density variables such as minimum area cover, minimum tree height, minimum crown-cover percent, or minimum width. And the other, with some special spatial features of the territory such as the presence of plantations, agricultural activities or non-forest trees within the forest itself \cite{klein91,klein92,klein2001,Lund2002,vidal2008}. The problem is that important ecosystem services as carbon sequestration may be lost when natural forests are severely degraded or replaced by plantations; but technically it remains as forests under these definitions. 

In this paper we will focus on a feature missing in most forest definitions, its complexity. A system is complex when it presents a sufficiently large number of elements that have strong enough interactions and/or its configuration space changes fast enough (in terms of the scales of the observer). Morin has suggested \cite{Morin2007} that all systems are complex, as there are not stationary or non-interactive systems. And certainly forest and forest management occupies a very high place in the gradient of complexity.

\begin{figure}
\begin{centering}
\label{figComplexDiag}\includegraphics[scale=0.45]{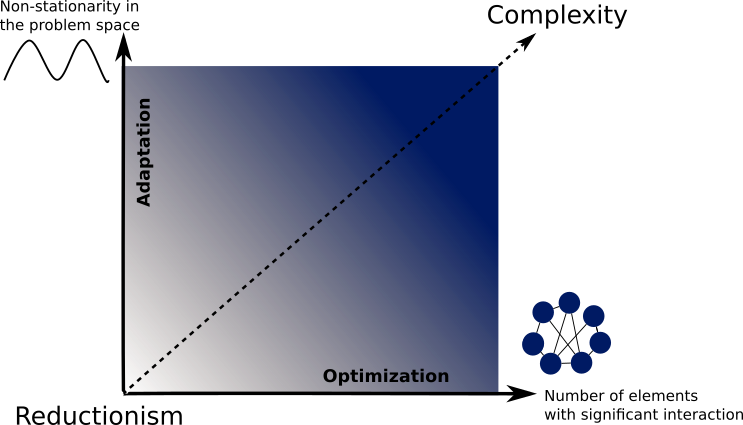}
\par\end{centering}
\caption{A system is complex when presents a sufficiently large number of elements that have strong enough interactions (the problem of the three bodies is already insoluble) and / or configuration space changes fast enough (in terms of scale-space observer ).}

\end{figure}

Already since the work of \cite{kimmins} and later by \cite{carey}, some components of forests complexity have been identified. The first aspect of complexity known as the alpha diversity, is the interaction complexity due to: (a) structure complexity that emerge from interactions going from the competition among trees of the same age and results in dominant, codominant, subordinate, and suppressed trees structure; interaction among trees of different ages, as trees eventually get damaged, affected by pests or breakdown they interact in distinct ways depending on their health state; interactions with the environment mainly due to the fact that variability in light, temperature, and soil moisture promotes structurally-diverse growth on the forest floor; and interactions related to canopy stratification where trees of different ages and growth habits produce multiple layers of vegetation; (b) species composition complexity, which comes mainly from habitat variability caused by patchy canopies that produce variability in light, temperature, and soil moisture; leading to patches of different types in the understory. On the other hand we have the pre-interactive niche diversification that in general terms affects the expansion in forest structure as well as plant species composition; and (c) functional complexity that emerges from the combined effect of structure and composition complexity, and is associated with high carrying capacities for diverse animals, high productivity for plants, effective regulation of nutrients and water cycling and ultimately defines the forest health and resilience capacity. 

In second terms we have the complexity that arises from the evolution of the configuration state-space, represented by the variability of the alpha diversity. The changes of alpha variability across local landscapes and across different climatic regions are the so called beta and gamma diversities respectively. Local landscape diversity or beta diversity is driven by the local topographic variations that lead to local gradients of soil moisture, depth, chemistry and physical properties in the soil and microclimates variations for example. The gamma diversity is associated with the ecosystem response to disturbances such as fire, insects, floods, pathogens, human activity, and the autogenic successional recovery of the disturbed ecosystems. Small time scale dynamics will affect mostly alpha diversity; while high time scale dynamic changes the beta diversity resulting for example in a shifting mosaic of different alpha-level ecosystem conditions \cite{kimmins, carey}.

In \cite{Gershenson2012,Fernandez2013} the authors use information theory to provide abstract and concise measures of complexity as well as other informational concepts associated with it, such as: emergence, self-organization, and homeostasis. Focusing on the information produced by a system, complexity represents the balance between emergence and self-organization, while homeostasis can be seen as a measure of the stability of the system. For details in this approximation see Prokopenko et al \cite{Prokopenko2011}.

Nevertheless, defining an specific measure for complexity is an open task that is being approached from different perspectives. Two recent interesting approaches are the informational framework and time series analysis related with more dynamic characteristics of the system.

Taking this into account, we present three case-studies in order to illustrate three relevant components of forest complexity and their implications.

\section{Methods}

\subsection{Systemic complexity in Zoquiapan Forest}

The first case corresponds to the Universidad Aut\'onoma Chapingo forestry experimental field of Zoquiapan (19,418ha) which was decreed as a Natural Protected Area (NPA) on March 13, 1937 by president Lazaro Cardenas \cite{LomasBarrie}. 

The park belongs to the physiographic region of Lagos y Volcanes del Anahuac province, in the Gran Sierra Volcanica subprovince with a predominance of andesita rocks \cite{INEGI81, Leet}. The site is characterized by seven units of soil, being  andosols the most abundant (29.81\%) and the most suitable for forestry\cite{Luna,Fistpratick,Consejo}.

With a dominant semi-cold climate in the highlands and template in the lowlands; the park falls within the watersheds of Texcoco-Zumpango Lake in the west, and in the north by the Tuchac lake and Tecocomulco \cite{spp}.

The park's vegetation is mainly characterized as the so-called temperate forests being the pine forest the most extensive, followed by the pine forests of \textit{P. hartwegii}, \textit{P. ayacahuite} and \textit{P. pseudostrobus}. For its part, the \textit{Quercus} forest is the second most representative vegetation in the park followed by grasslands, some of which are climax communities on the tops of the highest peaks, and some are the product of a disturbance in the original forest giving place to the so called anthropogenic grasslands, as well as those which are in the valleys where floods are periodic due to soil drainage or high precipitations \cite{RZEDOWSKI}.

In Table 1, we present some of the environmental system elements and potential variables to measure systemic complexity.

\begin{table}
\begin{centering}
\begin{tabular}{|c|c|}
\hline 
Concept & Variables\tabularnewline
\hline 
\hline 
Vegetation & Trees, shrubs and herbaceous species\tabularnewline
\hline 
Soil & Biotic and physiochemical characteristics\tabularnewline
\hline 
Weather & Temperature, precipitation, winds, humidity, evaporation \tabularnewline
\hline 
Light & Solar radiation, light hours per day\tabularnewline
\hline 
Wildlife & Reptiles, mammals, birds, etc\tabularnewline
\hline 
Water & Runoff, infiltration, evaporation, groundwater flux\tabularnewline
\hline 
Orography & Topography, altitude, slopes\tabularnewline
\hline 
\end{tabular}
\par\end{centering}
\caption{Some environmental system elements and possible variables for complexity measurement}

\end{table}

Considering existing information, we selected the variables that appear in Table 2 which are climatological data (59 years from 1951-2010) such as precipitation, temperature and evaporation taken from the  National Water Commission database; vegetation data; wildlife and soil.

\begin{table}
\begin{centering}
\begin{tabular}{cccc}
Variable & Minimum & Maximum & Mean\tabularnewline
\hline 
\hline 
Temperature ($^{\circ}$C) & 3 & 21 & 11.6\tabularnewline
Precipitation (mm) & 8 & 181 & 64.4\tabularnewline
Evaporation (mm) & 60 & 133.3 & 86.5\tabularnewline
Altitute (m.a.s.l.) & 3075 & 3679 & 3250\tabularnewline
Soil-pH & 6.8 & 7.1 & 6\tabularnewline
Soil-color & Red & Black & Black\tabularnewline
Terrain slope & 10 & 48 & 25\tabularnewline
Soil depth & 0 & 100 & slope dependent\tabularnewline
\end{tabular}
\par\end{centering}
\caption{Maximum, minimum and mean values for the environmental variables taken
into account.}
\end{table}

In order to measure the systemic complexity of the Zoaquiapan Forest we used the Gershenson and co-authors \cite{Gershenson2012,Fernandez2013} informational frame work based on Shannon information defined as

\begin{equation}
I=-\sum p(x)\log P(x)\label{eq:shannon}
\end{equation}

If we describe phenomena in terms of information, in order to have new information, old one has to be transformed. Thus, information emergence $E$ represents the rate of information transformation therefore, we can say that emergence is the same as Shannon's information $I$ . 

For its part, Self-organization ($S$) has been correlated with an increase in order, i.e. as a reduction of entropy \cite{Gershenson-Hey2003}. Then if emergence implies an increase of information, which is analogous to entropy, self-organization should be anti-correlated with emergence in such way that 

\begin{equation}
S=1-I=1-E.
\end{equation}

Now then, as complexity can be seen as the amount of information required to describe a phenomenon at a particular scale \cite{BarYam-2004}. Then complexity is an extensive variable which numerical value depends on the particular scale of analysis. Although there have been several ways for quantifying complexity \cite{GelMann1996,Langton1990,Wuensche1999}, because its relation with entropy of particular interest is the definition of \cite{Lopez-Ruiz1995}. Highly ordered systems (low information), such as crystals that are far from equilibrium, will be held a low complexity. Also, highly chaotic systems (high information), such as gases that are close to equilibrium, are held a low complexity. Then, high complexities are achieved with balanced values of information and disequilibrium.

Finally, a dynamical system has a high homeostatic capacity if it is able to maintain its dynamics close to a certain state or states (attractors). When perturbations or environmental changes occur, the system adapts to face the changes within the viability zone, i.e. without the system breaking, Homeostasis is also strongly related to robustness. 

In this way, following \cite{Gershenson2012,Fernandez2013,Lopez-Ruiz1995} complexity and other information concepts will be measure as

\begin{equation}
\begin{array}{ccc}
E & = & \frac{I_{out}}{I_{in}},\\
S & = & I_{in}-I_{out},\\
C & = & E*S,\\
H & = & 1-d(I_{in},I_{out}),
\end{array}
\end{equation}

where $I_{in}$ may be set to zero without loss of generalization, and then in particular measure of complexity may be  $C=a*I_{out}(1-I_{out}),$ with $a,d$ some normalizing constants. 

\subsection{Complexity from species and structure composition in Riparian vegetation}
In general, the term Riparian vegetation has been used to refer to plants that are set on the banks of streams, rivers, ponds, lakes and some wetlands \cite{Naiman-97}. Consequently, these areas are interfaces between aquatic and terrestrial systems, and encompass a fine gradient of the physical environment of natural communities \cite{Naiman-93,Naiman-97}. Riparian Forests provide us with essential ecological functions and have a strong influence on adjacent ecosystems, therefore its alteration has wide landscape implications \cite{Naimen-98}. Furthermore, among ecosystem services provided by riparian vegetation are: the protection of water quality as it protects the watershed filtering surface flow and subsurface adjacent areas, reducing runoff and promoting infiltration \cite{Guilian-94}; their role as biological corridors \cite{Harris-91}; biodiversity of habitats and species \cite{Naiman-93}, for example, it has been observed that these forests support a diversity and density of migratory and resident birds \cite{warkentin-95}; it also provides certain species, like scarlet macaw, with resources such as nesting and feeding sites \cite{renton-04}. Additionally, riparian vegetation plays an important role in stabilizing the soil, protecting it from the erosive forces of water \cite{Harmel-99}. It has also been observed that maintaining strips of riparian vegetation is a useful management practice to minimize the impacts of deforestation on aquatic systems, particularly to ensure the quality of water resource in our country is highly limited.

This case-study was conducted in a section of approximately 48 km on the banks of the Lacant\'un River, which adjoins the region known as Montes Azules Biosphere Protected Area. It is precisely in the evergreen tropical forests where there have been the fewest number of studies on riparian vegetation \cite{i=0000F1ogo-96}. 

The Lacant\'un River Basin is located within the hydrological region 30, called Grijalva-Usumacinta on the eastern side of Mexico, which covers the Lacant\'un River basin, the Chixoy River, the River Usumacinta and the Grijalva River \cite{INEGI-200a}.

\begin{figure}
\begin{centering}
\includegraphics[scale=0.45]{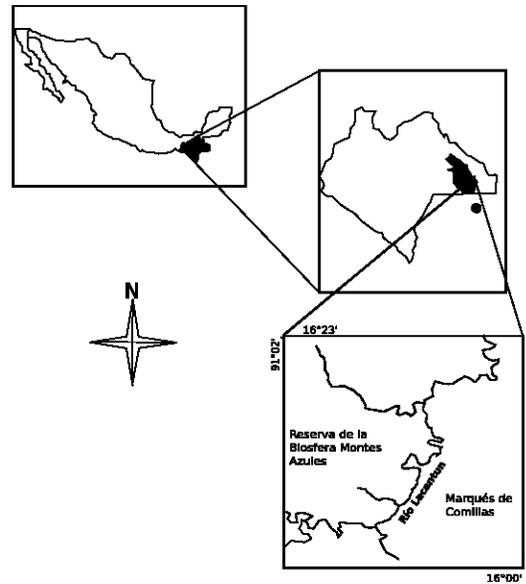}
\par\end{centering}
\caption{Location of the study area and view of a classification obtained from a Landsat ETM satellite image (2000). Green: conserved areas that maintain their vegetation cover, mostly high evergreen forest. Light yellow: areas that have lost their vegetation cover and have become mostly in fields and pastures of the suburbs.}

\end{figure}

The case-study area includes one section on Montes Azules and the other bank of the River Lacant\'un, which belongs to the region Marqu\'es de Comillas. These areas are characterized by vegetation that has been declining very fast in the last three decades due to deforestation, leaving an increasingly fragmented and less spatial continuity in the landscape. 

In Figure 4 we show conserved areas in green that keep their plant cover and in light yellow the areas that have lost their vegetation cover and have become mostly in to grass and cropland. In general, the figure shows that the left bank of the River Lacant\'un presents vegetation in better conditions; while on the right side a great fragmentation is observed.

To determine the vegetation diversity of riparian forest in the Lacant\'un river, we took 12 vegetation samples within an area of 1000 $m^{2}$ (0.1 ha) each. Census for each vegetation followed section of the river Lacant\'un for about 48 km, which allowed us to include natural variability of vegetation. The selection of the 12 sites for Vegetation surveys considered the presence of three different plant communities: 1) Five censuses preserved riparian vegetation, located on the banks within Montes Azules protected area. 2) Five censuses considered disturbed riparian vegetation, located on the banks of Marqu\'es de Comillas town. 3) Two censuses were made in the high evergreen forest area, which made it possible for us to obtain information from the original vegetation matrix. We used INEGI topographic maps (1:50 000) and a Landsat ETM satellite image 2000. 

\begin{figure}
\begin{centering}
\includegraphics[scale=0.45]{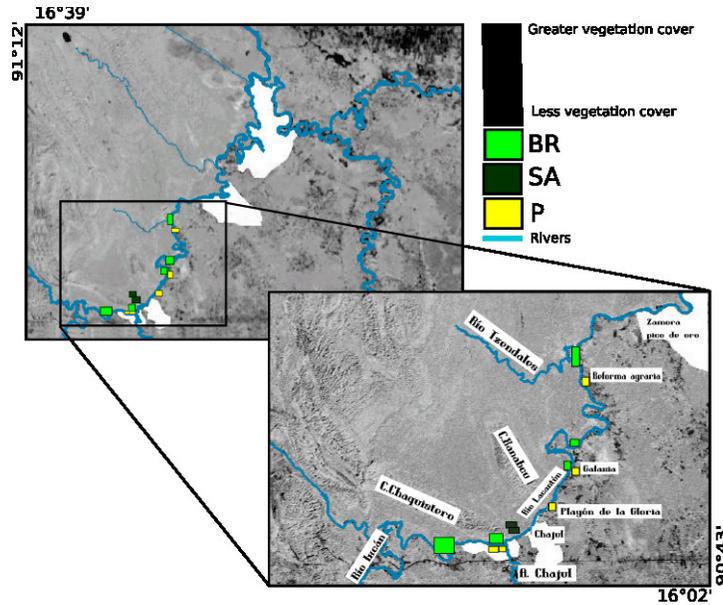}
\par\end{centering}
\caption{Location of the twelve sites along the Lacant\'un river over an image of the region showing a vegetation index obtained from a satellite image Landsat ETM (2000). Darker tones indicate greater vegetation cover, while lighter tones indicates less coverage. In the box above the colored boxes represent the differences in vegetation: RF = Riparian forest, J = jungle (evergreen forest area) and D = Disturbed vegetation}

\end{figure}

To quantify the plant diversity species locally, we used the Gentry method with some modifications \cite{Gentri-82,gentri-88,trejodirzo-02,trejodirzo-98}.

Plant diversity was measured using the Shannon index H which has the exact same functional form as the Shannon Information equation (Eq.\ref{eq:shannon}). In this case of study only comparison between vegetation complexity was made without taking into account other sources of complexity.

\subsection{Dynamic complexity using  time series analysis of the Metolius Forest,  Oregon, USA}

The last case-study is the The Metolius Forest which is located in the Cascade Range in Oregon, USA. This area is located approximately 15 miles northwest of Sisters, Oregon on the east slope of the Cascade Mountains. In the summer of 2003 it took place a series of wildfires called The B\&B Complex Fires that together burned 90,769 acres (367.33 $km^{2}$) of this Oregon forest. The fire complex began as two separate fires, the Bear Butte Fire and the Booth Fire; those eventually burned together, forming a single fire area that stretched along the crest of the Cascade Mountains between Mount Jefferson and Mount Washington. ( "B\&B Complex Forest and Fire History", Oregon Websites and Watersheds Project, Philomath, Oregon, January , 2011.; District, S. R. (2005). B\&B Fire recovery project record of decision.)

Interestingly,  this area has been monitored by the AmeriFlux project which is a network of PI-managed sites measuring ecosystem CO2, water, and energy fluxes in North, Central and South America. It was established to connect research on field sites representing major climate and ecological biomes, including tundra, grasslands, savanna, crops, and conifer, deciduous, and tropical forests. The network was launched in 1996, after an international workshop on flux measurements in La Thuile, Italy, in 1995, where some of the first year-long flux measurements were presented. The network grew from about 15 sites in 1997 to more than 110 active sites registered today (http://ameriflux.lbl.gov/ consulted on November 10, 2016).

In terms of dynamic complexity, it has been proposed \cite{LopezCorona2013} that $1/f$ noise is a fingerprint of a statistical phase transition, from white noise (randomness -- no correlation -- disorder) to brown noise (predictability -- correlation -- order), being also then a fingerprint of complexity, statement that is being supported by new evidence from several different disciplines as medicine \cite{Silva-2015,rivera-2016}, hydrology \cite{Sinngh-2015} and economy \cite{Niu-2015}.

Colored time series such as $1/f$ noise are part of a set of scale invariant signals defined by an inverse power law of the form 

\begin{equation}
S(f)\sim1/f^{\beta}
\end{equation}

where S is the Fourier Transformation of the signal, f is the frequency and $\beta$ is the spectral density exponent, which classifies the signals depending of its value: $\beta=0$ for white noise, $\beta=1$ for pink and $\beta=2$ for brown noise. 

These tree types of noise exhibit quite different statistical characteristics. The correlations are zero for white noise, large for brown noise, and infinite for a periodic series. Then the parameter $\beta$ gives a measure of the correlation strength and may be used as a control parameter for complexity \cite{LopezCorona2013}. 

In here we analyze the time series of $CO_{2}$ fluctuations (measured on a half an hour basis) from the Metiolus Pine Forest (Oregon, USA) for 2001. Data where collected from the described Ameriflux \cite{xiao-ameriflux} data base ({http://ameriflux.lbl.gov} and analyzed using two different methods to calculate the spectral density exponent $\beta$.  First we used the traditional Fourier analysis and afterwards a Detrended Fluctuation Analysis approach. 

Fourier Analysis is the study of the way general functions may be represented or approximated by sums of simpler trigonometric functions and is described in any general signal analysis textbook. For its part, Detrended Fluctuation Analysis (DFA) is a method for determining the statistical self-affinity of a signal. 

Suppose an accumulated series, 

\begin{equation}
Y(k)=\stackrel[i=1]{k}{\sum}u(i)-\bar{u},
\end{equation}

where $u(i)$ is the i-th value of the time series and $\bar{u}$ its mean value. Then consider that $Y(k)$ is divided into windows of size $n$ and for each window, we define a function representing the local trend fitted to the data and removed from the original values, resulting in a detrended accumulated time series $Y_{d}(k)$. Lets take

\begin{equation}
F(n)=\sqrt{\frac{1}{N}\stackrel[k=1]{N}{\sum}\left|Y_{d}(k)\right|^{2}}
\end{equation}
a function that represents the fluctuations within windows of size $n$. The procedure is usually repeated to increase window sizes in order to define the relationship between $F(n)$ and $n$. In general, $F(n)$ follows a power law with a scaling exponent (the Hurst exponent) related with the spectral density exponent $\beta$.

\section{Results}

\subsection{Zoquiapan Forest Complexity}

The results on the emergence, self-organization, complexity and homeostasis, calculated using Eq. 6, are summarized in Table 3. Also the scale for results evaluation is included in Table 4. 

\begin{table}
\begin{centering}
\begin{tabular}{ccccc}
 & E & S & C & H\tabularnewline
\hline 
\hline 
Vegetation & 0.268 & 0.732 & 0.785 & \tabularnewline
Wildlife & 0.242 & 0.758 & 0.734 & \tabularnewline
Soil & 0.133 & 0.867 & 0.460 & \tabularnewline
Temperature & 0.157 & 0.843 & 0.529 & 0.436\tabularnewline
Precipitation & 0.201 & 0.799 & 0.643 & 0.340\tabularnewline
Evaporation & 0.198 & 0.802 & 0.636 & 0.324\tabularnewline
\hline 
\textbf{EFEZ system} & \textbf{0.200} & \textbf{0.800} & \textbf{0.631} & \textbf{0.367}\tabularnewline
\end{tabular}
\par\end{centering}
\caption{Informational measurements for the Estaci\'on Forestal Experimental Zoquiapan system (EFEZ): Emergence (E), Self-organization (S), Complexity (C) and Homeostasis (H). Not sufficient data were available for calculating
H for all variables. }

\end{table}

From Figure1, the variables of soil, temperature, and evaporation are in a category of very low emergence; meanwhile vegetation, wildlife and precipitation are cataloged as low emergence. In terms of self-organization, the vegetation, wildlife and precipitation variables have a high self-organization value; just as the variables of soil temperature and evaporation which are also classified as high self-organization. It should be noted that when comparing the figure emergence and self-organization, the inverse relationship between them is clear.

We also notice that the variables on average have a high complexity value; being the soil and temperature in a category of medium complexity, compared with the other variables (vegetation, wildlife, precipitation
and evaporation) which have high complexity values range.

\begin{figure}
\begin{centering}
\includegraphics[angle=-90,scale=0.55]{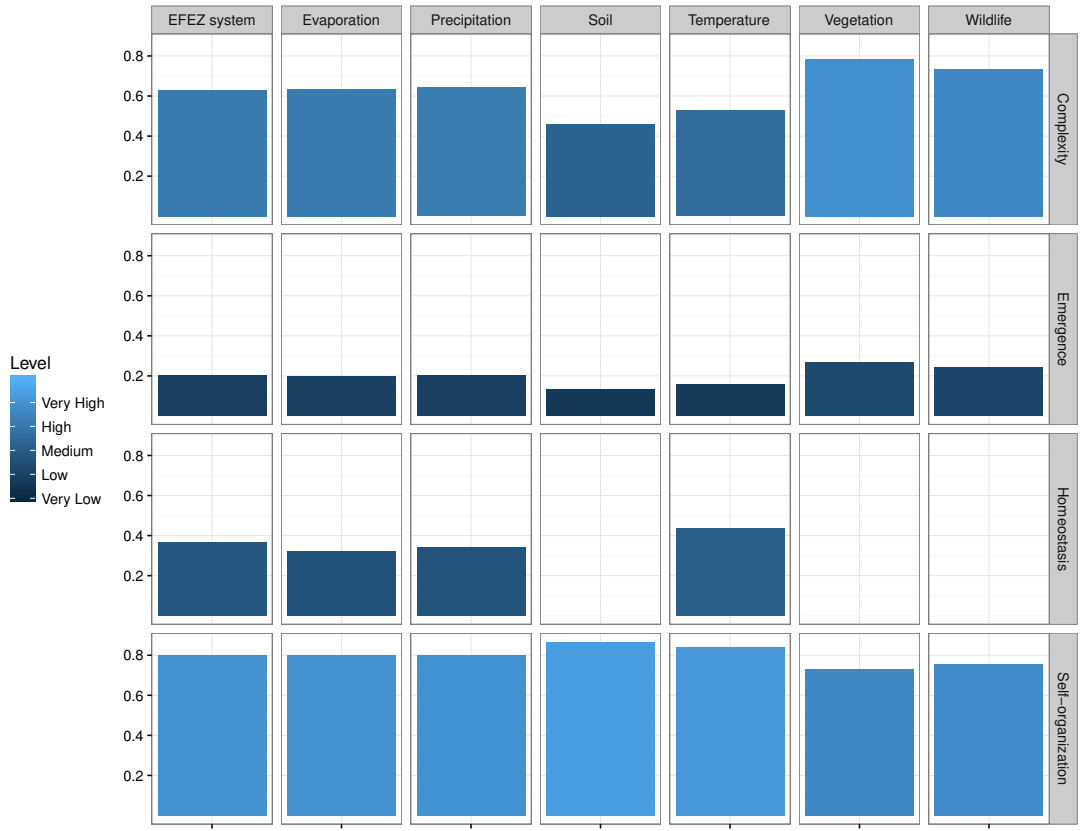}
\par\end{centering}
\caption{Results for informational measurements for complexity in Zoquiapan forest. Horizontal axis correspond to environmental variables and vertical axis shows the different informational quantities. Dark colors are related with low values of the informational, while light ones are related with high values. The correspondence between color and numeric informational values is as follows, Very High: 0.8-1, High: 0.6-0.8, Medium: 0.5-0.6, Low: 0.2-0.4, Very Low: 0-0.2. From results it can be summarize that this study-case in Zoquiapan forest is in high levels of Self-organization and Complexity, while it exhibits low levels of Emergence and depending on the variable from low to medium levels of Homeostasis }
\end{figure}

From the results above, it can be summarized that among the six variables analyzed, vegetation is the one with major emergency (0.26), and therefore less self-organization (0.73); conversely soils present the lowest emergency value (0.13) and hence the greatest self-organization level (0.86). Vegetation variable which is more complex (0.78) represents a better balance between emergency values and self-organization. Soil has the least complex variable (0.46), this may be because the scale of analysis incorporated very few soil variability in the site of study. In general, we observed that this temperate forest has a very low emergency value (0.20), while the degree of self-organization is very high (0.80) and system is also at high level of complexity (0.63).

\subsection{Comparing complexity in Lacant\'un river, is Riparian forest different?} 

In Figure 5, we present complexity over different type of vegetation (horizontal axis) and the individuals with DAP $>$ 1 cm and DAP $<$ 1 cm (vertical axis).

\begin{figure}
\begin{centering}
\includegraphics[angle=270,scale=0.55]{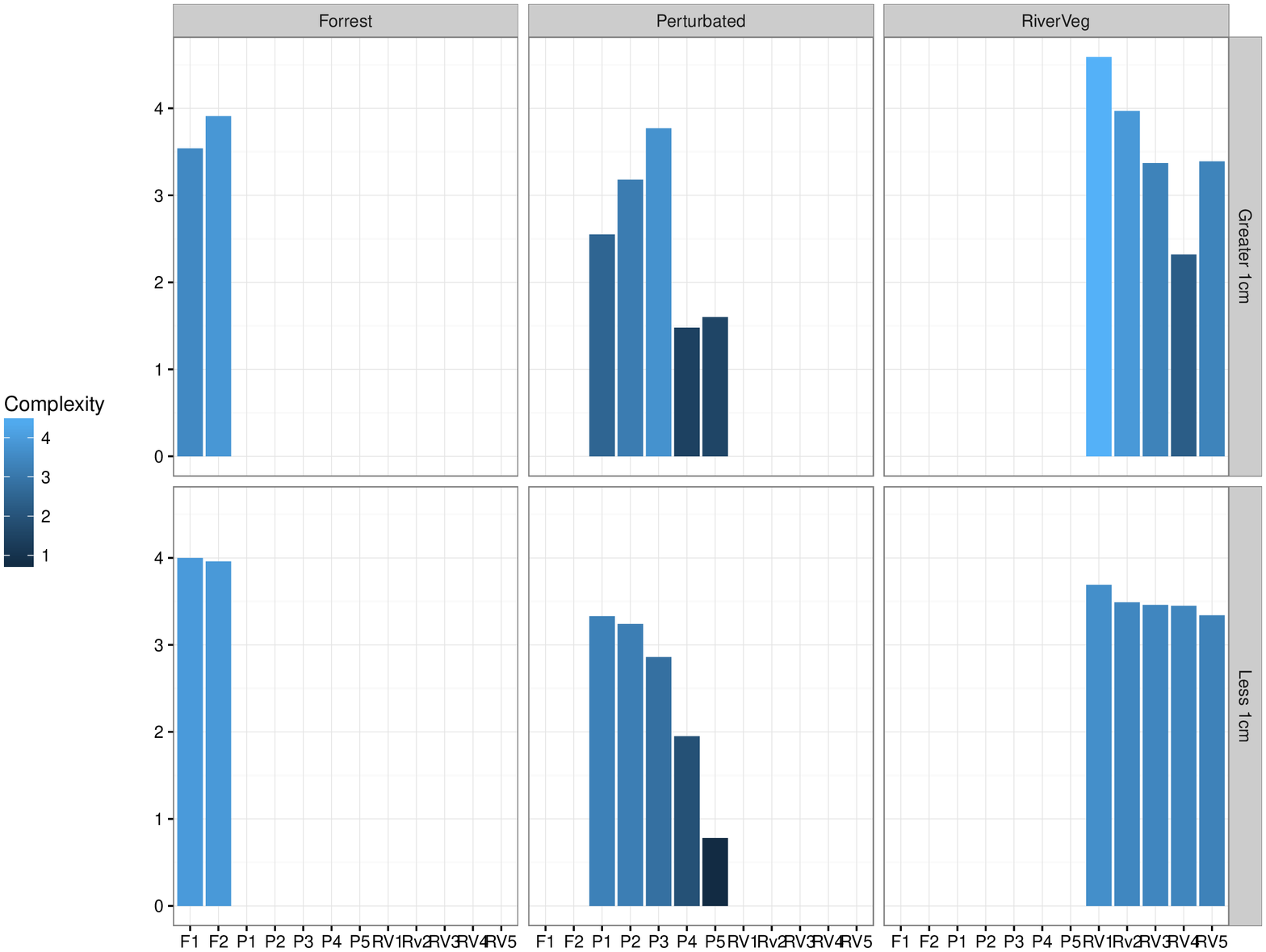}
\par\end{centering}
\caption{Complexity measurement over different type of vegetation and for the individuals with Diameter at breast height or DAP $>$ 1 cm and DAP $<$ 1 cm. As expected higher values of complexity corresponds to Riparian and Forest burn there are also some perturbed sites that remains complex due to the management }
\end{figure}

\begin{figure}
\begin{centering}
\includegraphics[angle=270,scale=0.35]{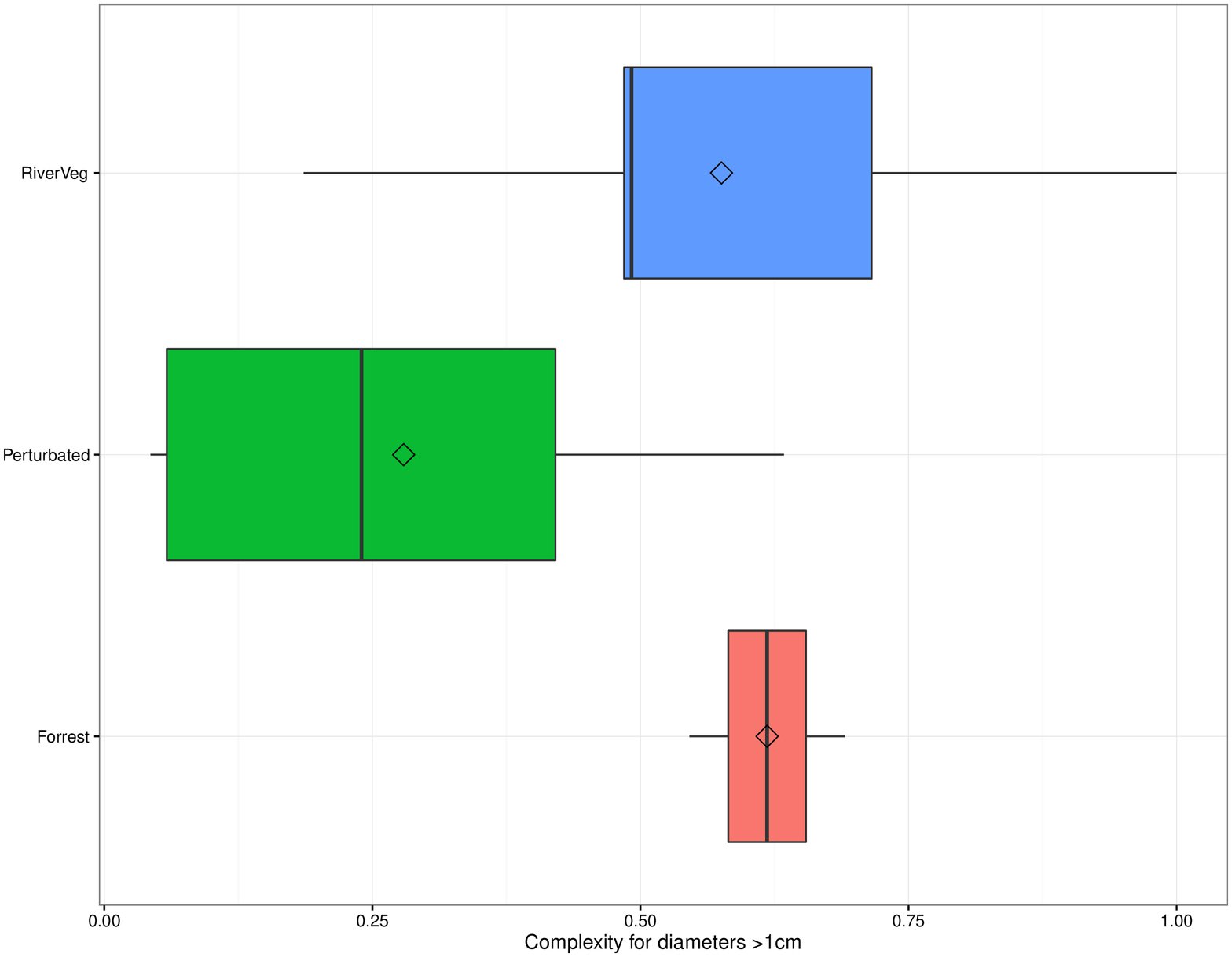}\qquad{}
\includegraphics[angle=270,scale=0.35]{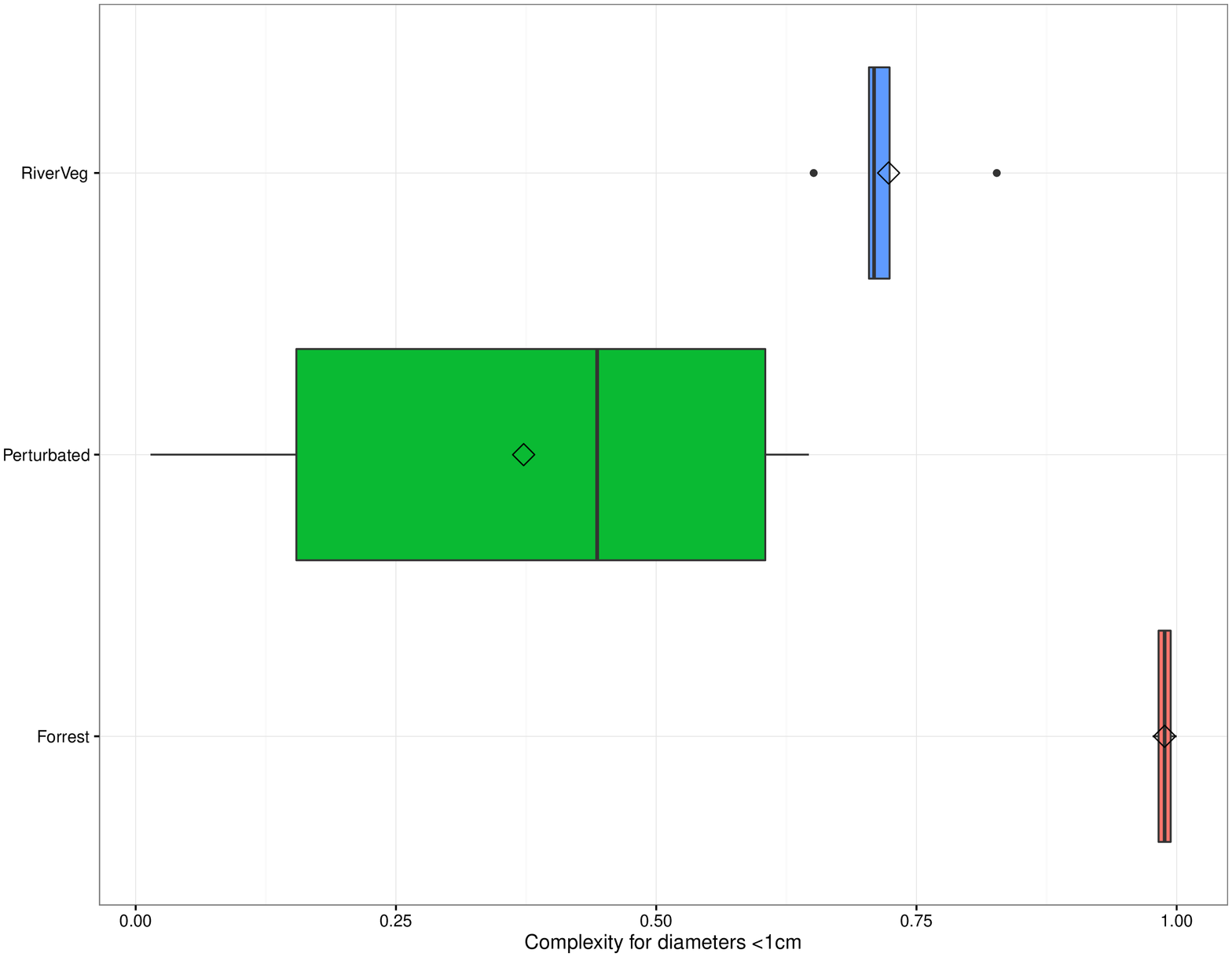}
\par\end{centering}
\caption{The top subfigure shows the complexity measurement boxplot over different types of vegetation and for the individuals with DAP $>$ 1cm, the diamond is the media of the population; on the bottom we present the complexity measurement boxplot over different type of vegetation and for the individuals with DAP $<$ 1cm, the diamond is the media of the population.}

\end{figure}

In general perturbed vegetation presents the lowest complexity values, while the highest complexity values are present in Forest juveniles (DAP $<$ 1 cm). Vegetation in Riparian forest (DAP $>$ 1 cm) show higher levels of complexity, which may be understood as a consequence of the regular and consistent geomorphological disturbances it experiences, such as continuous change in the topography of the canal. In fact, for some authors suggest that riparian forests are one of the most complex ecological systems of the biosphere and also one of the most important for maintaining the landscape and its rivers \cite{Naiman-97}.

It is also interesting that there are some outliers with high values of complexity in perturbed sites, as a result of good management practices which includes the conservation of several adult trees. Finally it is clear that the three types of vegetation exhibited different levels of complexity which are related to other informational concepts such as homeostasis (related to resilience) and even with sustainability potential \cite{LopezCorona2015-soc}.

\subsection{Forest's Dynamic Complexity}

 The time series analysis of $CO_{2}$ fluctuations from Metolius Pine Forest in Oregon, using  both Fourier and DFA, results in a $1/f$ behavior which is a fingerprint of complexity.

\begin{figure}[ht]
\begin{centering}
\includegraphics[scale=0.5]{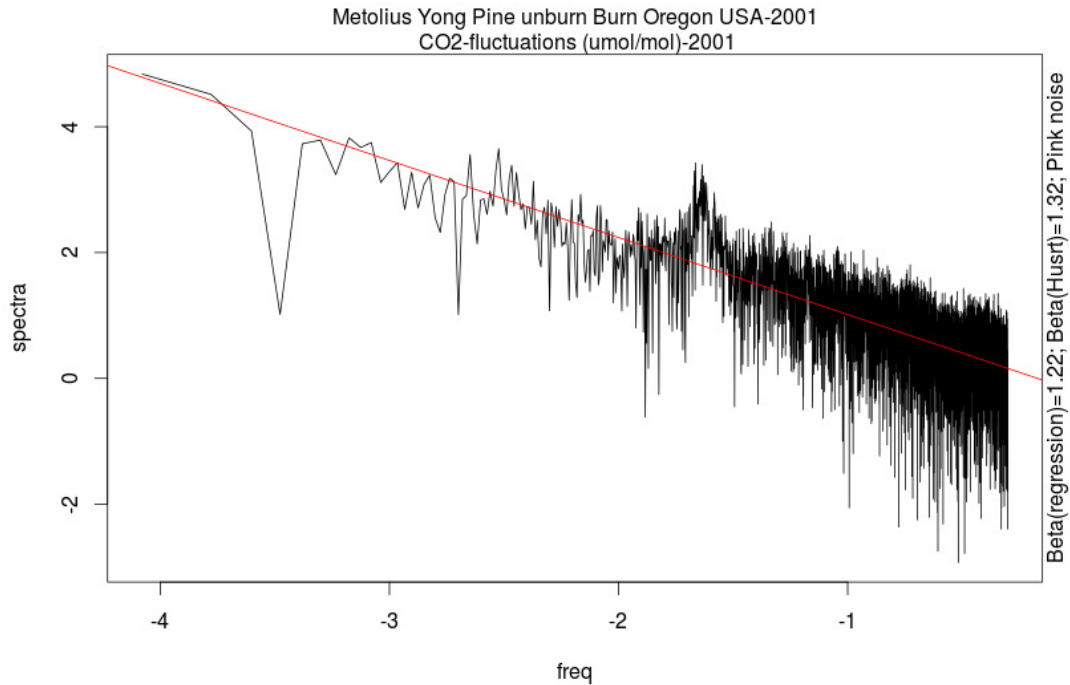}
\par\end{centering}
\caption{Power spectrum of CO2 Flux fluctuations at Metolius Pine Forest in Oregon, USA for the 2001 year. The signal is consistent with a Pink noise type which is related with dynamic complexity (L\'opez-Corona et al, 2013)}
\end{figure}

\section{Discussion}

In this paper we have showed different aspects of complexity: systemic complexity using the Gershenson formalism in Zoquiapan forest; biological complexity in the tropical rain forest and dynamic complexity in a Oregon forest (USA). From these cases of study we state the following: (1) from a systemic perspective forests exhibit high levels of complexity and very high levels of self-organization, which is consistent with previous work \cite{Boyer}; (2) from our case of study in Lacantun river we showed that lowest values of complexity correspond to perturbed sites. However, this lost of complexity  may be reduce with a good management; (3) as suspected, riparian forest are different from tropical forest and in general for DAP $>$ 1 cm shows higher levels of complexity; (4) $CO_{2}$ fluctuation time series from Metolius Pine Forest in Oregon USA, follows a complex dynamic characterized by $1/f$ noise. Interesting enough, analyzing data from after the B\&B fires, the Metolius Forest lost complexity exhibiting a Brown like noise with beta above 1.5 which suggests a very interesting application of forest complexity measurements in terms of ecosystem integrity. New results that will be presented in other paper, show that most of North America Forest in AMERIFLUX data base also exhibits $1/f$ dynamics, which was expected because it represents a good evolutionary trade-off between robustness and adaptation. 

Recently \cite{LopezCorona2015-soc} and co-workers have pointed out
that as complexity is deeply related with sustainability. This fact
has great implication for forest management and conservation, for example in the Lacantun case of study it is represented with the presence of high complexity outliers in perturbed sites. We know first hand, thanks to the field trips to the sites and the interaction with the stakeholders, that this complexity values correspond to an agricultural site with a very good management that includes the conservation of several adult trees. This proves that under a proper management we can conserve systemic properties of the agro-ecological system such as complexity, promoting sustainability. Another example, but in a negative direction, is the  common practice of reforestation that uses large mono-species plantations, this of course, has a much lower complexity value than original forests. Then even when reforestation is a good practice, our results make it clear that it should be done in such a way that it maximizes complexity (and sustainability), being biodiversity the most straightforward strategy to achieve so. In the same manner, forest engineers cut some trees in the so called sanitation cutting, trying to increase forest productivity by liberating resources eliminating old, damage or sick trees. But once again this practice does not take into account that diversity (complexity) is profoundly necessary to maintain forest sustainability. 

As forests are indeed complex systems, they are inherently non-predictable and non-controllable \cite{sang-2015} which causes very serious issues regarding the validity of traditional management plans and environmental impact assessment.
Management should be then re-conceptualized as a multi-objective optimization process with several restrictions over a non-stationary space of configuration. Therefore, traditional management should be replaced with an adaptive scheme. For its part, Environmental Impact Assessments (EIA) are intended to be technical and administrative instruments used to identify, prevent and interpret the environmental impacts that may occur in the environment if a project is executed, in the context of authority decision about the project acceptance. Based on our results which indicate that forest are so complex that any EIA is limited, we suggest that this may be partially corrected by incorporating a stochastic perspective in the EIA formulation, complemented with a final risk analysis.

\section*{Acknowledges}

We acknowledge the following AmeriFlux sites for their data records:
site IDs. In addition, funding for AmeriFlux data resources was provided by the U.S. Department of Energy's Office of Science. OLC thanks Fondo Capital Semilla at Universidad Iberoamericana and to SNI program with number 62929. ERM thanks CONACyT Posdoctoral Fellowship.

\end{document}